\documentclass[final]{IEEEtran} 
\usepackage{amssymb,amsmath,amsthm,bm}
\usepackage{graphicx,color}
\usepackage{subfigure}
\graphicspath{{figures/}}
\usepackage{cite}
\usepackage{url}
\usepackage{enumerate}
\usepackage{pgfplots}

\usepackage{titlesec}
\titlespacing{\section}{0pt}{5pt}{2pt}
\titlespacing{\subsection}{0pt}{3pt}{2pt}
\titlespacing{\subsubsection}{0pt}{1.5pt}{1pt}

\usepackage[mathlines]{lineno}

\usepackage{multirow}
\usepackage{setspace}

\newlength\imagewidth
\newlength\imagewidtha
\newlength\imagewidthb
\newlength\figwidth
\newlength\figwidtha
\newlength\figwidthb
\newlength\figsep
\newtheorem{Theorem}{Theorem}

\newtheorem{Definition}{Definition}
\newtheorem{Property}{Property}

\DeclareMathOperator\supp{supp}
\let\vec\relax
\DeclareMathOperator\vec{vec}

\let\Re\relax
\DeclareMathOperator\Re{Re}

\let\Im\relax
\DeclareMathOperator\Im{Im}

\newcommand{\diag}{\mathop{\mathrm{diag}}}

\begin{document}

\title{{{Bi-level Protected Compressive Sampling}} 
\thanks{}
}

\author{Leo Yu Zhang,~\IEEEmembership{Student Member,~IEEE},
Kwok-Wo Wong,~\IEEEmembership{Senior Member,~IEEE}, \\
Yushu Zhang,
{Jiantao Zhou,~\IEEEmembership{Member,~IEEE},}

\thanks{Leo Yu Zhang and Kwok-wo Wong are with Department of Electronic Engineering, City University of Hong Kong, Hong Kong (e-mail: leocityu@gmail.com; itkwwong@cityu.edu.hk)}
\thanks{Yushu Zhang is with
the School of Electronics and Information Engineering, Southwest University, Chongqing 400715, China (e-mail: yushuboshi@163.com)}
\thanks{{Jiantao Zhou is with
Department of Computer and Information Science, Faculty of Science and Technology, University of Macau, Macau (e-mail: jtzhou@umac.mo)}}
}

\maketitle

\begin{abstract}
Some pioneering works have investigated embedding cryptographic properties in compressive sampling
(CS) in a way similar to one-time pad symmetric cipher.
This paper tackles the problem of constructing a CS-based symmetric cipher under the key reuse circumstance,
i.e., the cipher is resistant to common attacks even a fixed measurement matrix is used {multiple} times.
To this end, we suggest a bi-level protected CS (BLP-CS) model {which makes use of} the advantage of the {non-RIP measurement matrix construction}.
Specifically, two kinds of artificial basis mismatch techniques are investigated to construct key-related sparsifying bases.
It is demonstrated that the encoding process of BLP-CS is simply a random linear projection, which is the same as the basic CS model.
However, decoding the linear measurements requires knowledge of both the key-dependent sensing matrix and its sparsifying basis.
{The proposed model} is exemplified by sampling images {as a joint data acquisition and protection layer for resource-limited wireless sensors.
Simulation results and numerical analyses have justified that the new model can be applied in circumstances where the measurement matrix can be re-used.}

\begin{keywords}
compressive sampling, restricted isometry property, encryption, known/chosen-plaintext attack,  random projection.
\end{keywords}
\end{abstract}

\section{Introduction}
\label{sec:intro}
Compressive sampling (CS) has received extensive research attention in the last decade \cite{Donoho:CSIntro:TIT06,Candes:IntroCS:SPM08, Richard:CSIntro:SPM07}.
By utilizing the fact that natural signals are either sparse or compressible,
the CS theory demonstrates that such signals can be faithfully recovered from
a small set of linear, nonadaptive measurements, allowing sampling at a rate lower than that required by
the Nyquist-Shannon sampling theorem.


The use of CS for security purposes was first outlined in one of the foundation papers \cite{Candes:CS:TIT06}, in which Candes and Tao suggested that the measurement vector obtained from random
{subspace} linear projection
can be treated as ciphertext since the {unauthorized user} would not be able to decode it unless he knows in which random subspace the coefficients are expressed.
In this way, the entire CS scheme can be considered as a variant of
symmetric cipher, {where the signal to be sampled, the measurement vector and the measurement matrix are treated as the plaintext, the ciphertext and the secret key, respectively}.

{It is a favorable characteristic that certain kind of data protection mechanism can be embedded into the data acquisition stage. Such a property of CS is of particular importance for data acquisition systems in sensor networks, where each sensor is usually resource-limited and a separate cryptographic layer is too expensive for secure data transmission. Example applications work under this circumstance include visual sensor networks \cite{winkler2014security}, video surveillance networks \cite{dufaux2008scrambling} and etc. Meanwhile, CS paradigm also found to be useful for medical systems, especially in the case that sampling speed \cite{lustig2007sparse} and privacy \cite{barrows1996privacy} are two major concerns.}

{There are a} number of studies exploring the security that {a CS-based symmetric cipher} can provide from the computation point of view.
For example, it was shown in \cite{Rachlin:secrecy:08} that the measurement matrix leads to computational secrecy
under some attack scenarios, such as brute-force attack and ciphertext only attack (COA).
Based on this result, there were many attempts in establishing secure measurement matrices.
In \cite{dautov:establishing:2013}, constructing the measurement matrix using physical layer properties
and linear feedback shift register (LFSR) with the corresponding $m$-sequence was proposed.
{In \cite{tong2011compressive}, Tong \textit{et al.} suggested constructing CS measurement matrix by chaotic sequence for privacy protection in video sequence.
In \cite{Cambareri:TwoClassCS:2013}, Cambareri \textit{et al.} employed CS
to provide two access levels by artificially carrying out sign flips to a subset of the measurement matrix.
In this way, the first-class decoder, who can access full knowledge of the measurement matrix, can retrieve the signal faithfully while the second-class decoder, who can only access partial knowledge of the measurement matrix, subjects to a quality degradation during reconstruction.
The work was later extended to multi-class low-complexity CS-based encryption \cite{cambareri2015low}.}



{Another research area of the secrecy of CS lies in the information theory frame. It is shown in \cite{yangsecurity} that CS-based cryptosystems fail to satisfy both Shannon's
and Wyner's perfect Secrecy. In this context, Cambareri \textit{et al.} \cite{cambareri2015low}
defined an achievable security metric, i.e., asymptotic  spherical security, for CS-based cipher. Basically, it states that the statistical properties of the random measurements only leak information about the plaintexts' energy. Based on this observation, Bianchi \textit{et al.} \cite{Bianchi:SecRLM:ICASSP14} suggested that re-normalizing every measurement vector and treating the normalized measurements as the ciphertext will lead to a perfect ``securized" CS-based cipher with the help of an auxiliary secure channel to transmit the energy of the real measurement vector.}

{
It should be noted that all the above security features of CS-based ciphers are obtained under limited attack models, i.e., the adversary is permitted to work out the secret key or plaintext from ciphertext only or to search the entire key space. Under more threatening scenarios, such as known-plaintext attack (KPA) and chosen-plaintext attack (CPA), the adversary can easily reveal the measurement matrix (secret key in a CS-based cipher) if he is able to collect sufficient amount of independent plaintexts. As such, to maintain their respective security features,
all the results mentioned above must work in a one-time-sampling (OTS) manner, i.e., the measurement matrix is never re-used.
}

{
Assume that a $K\times M$ measurement matrix is produced by using a secure deterministic random number generator (SDRNG) from a secret key shared between the encoder and decoder.
We note that this is exactly the case of the traditional one-time-pad (OTP) cipher \cite{shannon1949communication}.
If a sparse signal belongs to $\{0, 1\}^{M}$, it requires exactly $M$ bits to perfectly protect this signal when OTP cipher is applied. For the case of OTS, it requires at least $K\times M$ bits (if the Bernoulli matrix is used) to sample (encrypt) the signal.
From this sense, the OTS CS-based cipher indeed reduces the service life of the SDRNG.
Meanwhile, generating a different measurement matrix for every signal could be energy-consuming.
Additionally, for engineering practice, using the same measurement matrix for multiple signals or signal segments flavors the subsequent source coding stage
of multimedia data sensing, as discussed in \cite{mun2012dpcm,liu2014joint}.
Based on these observations, it is concluded that investigating the behavior of CS-based cipher under the multi-time-sampling (MTS) scenario is both important
from the cryptographic and engineering point of view.
}

{
The work presented in \cite{cambareriknown} offers an intimate view for MTS CS-based cipher, where
a second-class user in the two-class CS encryption \cite{cambareri2015low} tries to upgrade the recovery
quality by studying only one pair of known-plaintext and ciphertext. Restricting the measurement matrix to the form of Bernoulli matrix, it is shown in
\cite{cambareriknown} that the number of candidate measurement matrices matching a single pair of known plaintext and ciphertext is too huge
for the adversary to search for the true one. Still, the result only holds for a single plaintext-ciphertext pair while in typical KPA the adversary can
access a large amount of plaintexts and the corresponding ciphertexts. Thus, the true measurement matrix may be determined uniquely.
The same argument also applies to the case of CPA.
}

{
A straight forward solution to support the usage of CS in MTS scenario is to encrypt the entire or only the significant part of the quantized measurement vector
using some conventional cryptographic method, such as AES or RSA.
However, as we mentioned earlier, a standalone encryption layer can be too costly for a CS sensor and this approach does not take advantage of the confidentiality provided by CS itself.
}

{
Another approach to achieve this goal is to embed other efficient cryptographic primitives in the the CS encoding process.
This is exactly the idea of product cipher introduced by Shannon \cite{shannon1949communication}, who suggested combining two or more cryptographic
primitives together such that the product is more secure than individual component against cryptanalysis.}

{
In \cite{zeng2012scrambling}, Zeng \textit{et al.} proposed a speech encryption algorithm by scrambling the CS measurements.
A similar idea was later applied for secure remote image sensing \cite{huang2015compression}.
For the purpose of image acquisition and confidentiality, Zhang \textit{et al.} \cite{yushu:SCS:14arxiv} suggested scrambling the frequency
coefficients before the CS encoding instead of scrambling the CS samples. Note that scrambling the frequency coefficients is a mature technique for multimedia
confidentiality in traditional coding system \cite{Wenjun:FrequencyScrambling:TM03}, the main advantage of employing this technique in the CS paradigm is that
a so-called ``acceptable" permutation can make the column (or row) sparsity level of $2$D signals uniform \cite{Fang:permutation:TSP13},
thus relaxing the restricted isometry property (RIP) of the measurement matrix and flavoring a parallel CS (PCS) reconstruction model.
The same technique is also used for privacy protection in cloud-assisted image service \cite{wu2014low}.
Another popular approach to form product cipher for MTS usage of compressive imaging is to employ
an optical encryption primitive, i.e., double-random phase encoding (DRPE) technique, such as those proposed in
\cite{deepan2014multiple,rawat2015compressive,li2015compressive}.
There is also work that try to embed low-complexity nonlinear diffusion into the measurements quantization stage to enhance security of CS-based cipher \cite{leozhang2015joint}.
}

{
Although the above mentioned product ciphers are efficient, generally they cannot resist CPA in MTS
scenario (this issue will be discussed in detail in Sec.~\ref{subsec:B} and \ref{subsec:C}).
The reason for the difficulty in applying CS-based cipher for MTS usage is due to the characteristic of
CS itself: 1) the signal to be sensed must be sparse; 2) the encoding process is linear.
For this reason, embedding some high-security primitives before CS encoding will probably make the signal noise-like
and not sparse anymore. On the other hand, the introduction of any non-linear cryptographic primitive in CS paradigm
will break the linearity of the sampling process and make the recovery infeasible.
}

{
Our work moves one step further for the usage of CS-based cipher under MTS scenario. Start with a RIPless
reconstruction observation, we study how to embed security features in sparsifying bases under the sparse constraint.
In more detail, we suggest a bi-level protected CS (BLP-CS) framework, which can be viewed as a product cipher of the basic CS model and transform-domain encryption technique under the sparse constraint.
In particular, we propose several techniques to construct secret key-related sparsifying basis and incorporate
them into our BLP-CS model. At the encoding stage, this model can be viewed as a new design of the measurement matrix,
thus the encoding is the same as that of the original CS model. However, a successful decoding requires knowledge of
the key-dependent sensing matrix and key-related sparsifying basis. In this way, the new product cipher can resist CPA.
}

{
This paper makes two contributions in the area of embedding secrecy in CS. On the one hand, we propose a CPA-resistant
product cipher by utilizing the confidentiality provided by CS. To the best of our knowledge, this is the first reprot that the CS-based (product)
cipher can resist CPA. On the other hand, we incorporate a cryptographic permutation to the CS encoding stage, thus relaxing the RIP of the
measurement matrix and flavoring a PCS reconstruction for $2$D sparse signals.
In this sense, our work can be considered as an extension of the work presented in \cite{Fang:permutation:TSP13}.
}


The rest of this paper is organized as follows. In Sec.~\ref{sec:II}, we first review the CS framework and present {the CPA on CS-based product ciphers}.
In Sec.~\ref{sec:secIII}, two techniques for constructing secret key-related sparsifying basis are proposed
to establish the bi-level protection model.
{Sec.~\ref{sec:discussion} presents comparisons of the OTS CS-based cipher and our BLP-CS model from complexity and security point of view.
}
As an application example, the new model is used to sample digital images in Sec.~\ref{sec:Sec4}. 
The superiority of the new CS-based image cipher is justified by both theoretical analyses and simulation results.
Our work is concluded in Sec.~\ref{sec:conclusion}.

\section{{Security Defects of Existing CS-based Ciphers in MTS Scenario}}
\label{sec:II}

{
As we mentioned earlier, there exists some effort to support CS-based cipher for MTS usage \cite{zeng2012scrambling,huang2015compression,yushu:SCS:14arxiv,wu2014low,deepan2014multiple,rawat2015compressive,li2015compressive}.
In this section, we report the fact that all of them fail to resist CPA. To begin with, we briefly review the theory of compressive sampling.
}

\subsection{CS Preliminaries}
We denote a $1$D discrete signal to be sampled as a column vector $\mathbf{x} = (x_1, x_2, \cdots, x_M)^T $. {\
$2$D signals of size $M = n \times n$, $\mathbf{X} = \left[\mathbf{X}_{i,j}\right]_{i=1, j=1}^{n, n}$, can be vectorized to $1$D format as $\mathbf{x}$ by stacking the columns of $\mathbf{X}$,
i.e., $\mathbf{x} = \vec(\mathbf{X})$.
}
$\mathbf{x}$ is said to be $k$-sparse under $\mathbf{\Psi}$
if there exists a certain sparsifying basis $\mathbf{\Psi} =\{\psi_{i,j}\}_{i=1,j=1}^{M,M}$ such that $\mathbf{x} = \mathbf{\Psi}\mathbf{s}$ and $\|\mathbf{s}\|_0=\# \{\supp{\mathbf{s}}\}=\#\{i:s_i \neq 0\} =k<<M$.
{
Here, we emphasize that in almost all of the works about the secrecy of CS, such as \cite{Rachlin:secrecy:08,cambareri2015low,Bianchi:SecRLM:ICASSP14,cambareriknown,zeng2012scrambling,li2015compressive}, the role of the basis is ignored or simply treated as an orthnormal matrix. We relax the requirement of the basis to an invertible matrix in this work.
The encoding process during CS is a linear projection, i.e.,}
\begin{equation}
\mathbf{y} = \mathbf{\Phi}\mathbf{x} = \mathbf{\Phi}\mathbf{\Psi} \mathbf{s} =\mathbf{A}\mathbf{s},
\label{eq:samplingspace}
\end{equation}
{if the sampling is perform in the space/time domain, or equivalently
\begin{equation}
\mathbf{y} = \mathbf{\Phi}\mathbf{s} = \mathbf{\Phi}\mathbf{\Psi}^{-1} \mathbf{x} =\mathbf{A}\mathbf{s},
\label{eq:samplingfrequency}
\end{equation}
if the sampling is performed in the frequency domain.}

{The revolutionary finding of CS is that the $K$ dimensional measurement vector $\mathbf{y}$  reserves all the information required for unique and stable recovery of $\mathbf{x}$ even if $k<K \ll M$ provided that the measurement matrix $\mathbf{A}$ obeys some information-preserving guarantees \cite{Candes:CS:TIT06,Richard:ProveRIP:2008,candes2011probabilistic,kueng2014ripless}. Since the linear systems (\ref{eq:samplingspace}) and (\ref{eq:samplingfrequency}) are undetermined, both of them have infinite solutions. Considering the signal is sparse, the} intuitive way to restore $\mathbf{x}$ is to solve the $l_0$ optimization problem
\begin{equation}
\label{eq:l0hard}
  \min \|\mathbf{{s}}\|_0~~\text{subject to }\mathbf{y}=\mathbf{A}  \mathbf{{s}},
\end{equation}
to obtain $\mathbf{{s}}$ and then recover $\mathbf{x}$ by  $\mathbf{x} = \mathbf{\Psi} \mathbf{{s}}$.
As stated in \cite{Tao:BPunderRIP:TIT05}, solving this problem is NP-hard because it
requires an exhaustive search over all subsets of columns of $\mathbf{A}$.

The convex relaxed form of problem~(\ref{eq:l0hard}) can be expressed as
\begin{equation}
\label{eq:l1opt}
  \min \|\mathbf{{s}}\|_1~~\text{subject to }\mathbf{y} =  \mathbf{A}  \mathbf{{s}}.
\end{equation}
As proved in \cite{Candes:CS:TIT06}, the solution of the $l_1$ problem~(\ref{eq:l1opt}) is identical to that of (\ref{eq:l0hard}) with overwhelming probability provided that $\mathbf{A}$ satisfies RIP. {Examples of widely accepted matrices satisfying RIP including Gaussian ensemble and Bernoulli ensemble with $K = O(k \log M)$ rows. Up to a logarithmic factor, the number of measurements is optimal \cite{Candes:CS:TIT06}. Here we note that all the previously mentioned approaches of embedding secrecy into CS-based (product) ciphers work with RIP.}
\begin{Definition} \cite{Richard:ProveRIP:2008}
\label{def:RIP}
A matrix $\mathbf{A}$ of size $K \times M$ is said to satisfy the restricted isometry property of order $k$ if there exists a constant $\delta_k \in (0,1)$ such that
\begin{equation*}
(1- \delta_k) \|\mathbf{x}^{(T)}\|_2^2 \leq \|\mathbf{A}^{(T)} \mathbf{x}^{(T)}\|_2^2 \leq (1+ \delta_k) \|\mathbf{x}^{(T)}\|_2^2
\end{equation*}
holds for all column indices sets $T$ with $\#{T}<k$, where $\mathbf{A}^{(T)}$ is a $K \times \#{T}$ matrix composed of the columns indexed
by $T$, $\mathbf{x}^{(T)}$ is a vector obtained by retaining only the entries indexed by $T$
and $\|\cdot\|_2$ denotes the $l_2$ norm of a vector.
\end{Definition}


{More generally, let the $K$ rows of $\mathbf{A}$, i.e., $\mathbf{a}_1^T, \cdots, \mathbf{a}_K^T$,  be i.i.d. random vectors drawn from a distribution, say $F$. The recently developed RIPless CS  theory states that the solution of problem~(\ref{eq:l1opt}) is unique and equal to that of problem~(\ref{eq:l0hard}) if the number of measurements grows proportionally to the product of coherence parameter and the condtion number of the covariance matrix \cite{candes2011probabilistic,kueng2014ripless}, as given by Theorem~\ref{theorem:RIPless}.
}
{
\begin{Theorem} \cite{kueng2014ripless}
\label{theorem:RIPless}
Let $\mathbf{s}$ be a $k$-sparse vector and $\omega \geq 1$. The solution of problem~(\ref{eq:l1opt}) is unique and equal to that of problem~(\ref{eq:l0hard}) with probability at least $1-e^{-\omega}$
if the number of measurements fulfills
\begin{equation*}
K = O(\mu(F)  \theta \cdot \omega^2 k \log M),
\end{equation*}
where $\mu$, the coherence parameter, is the smallest number that
\begin{equation*}
\max_{1\leq i \leq M } |<\mathbf{a}^T, \mathbf{e}_i>| \leq \mu(F)
\end{equation*}
and $\theta$ is the condition number of the covariance matrix $\Sigma= \mathbb{E}[\mathbf{a}\mathbf{a}^T]^{1/2}$ with $\mathbf{a}^T$ being a generic row random vector draw from $F$ and $\mathbf{e}_i$ being the canonical basis vector of dimension $M$.
\end{Theorem}}
{
What concerns us about the RIP CS and RIPless CS is that the quantity $\mu(F) \theta$ that governs the number of required measurements for successful $l_1$ reconstruction is different. For Gaussian, Bernoulli and partial Fourier matrices,
it is shown that $\mu(F) \theta = O(1)$ in \cite{candes2011probabilistic}. Moreover, it is easy to find out
that $\theta =1$ for unitary matrix and $\theta >1$ for generic matrix\footnote{Recall that condition number is the absolute value of the ratio between the largest and smallest singular values.}.
Moreover, the larger the value of $\mu(F)\theta$, the more the samples we need for exact reconstruction in the RIPless setting. We make us of this fact to design the measurement matrix for security purpose.}

{
In the subsequent sections, we will show that almost all the CS-based product ciphers mentioned above, i.e., those proposed in
\cite{zeng2012scrambling,huang2015compression,yushu:SCS:14arxiv,wu2014low,deepan2014multiple,rawat2015compressive,li2015compressive}, fail to resist the CPA under MTS scenario due to the fact that these product ciphers work only under the RIP framework.
}

\subsection{{Scrambling in the Measurements Domain or the Frequency Domain}}
\label{subsec:B}
{As described in the previous sections,  it is more practical if the same measurement matrix can be re-used multiple times.
To this end, there are some attempts trying to incorporate other low-complexity cryptographic primitives to fix the intrinsic security defect of CS in a manner of constructing product ciphers \cite{zeng2012scrambling,huang2015compression,yushu:SCS:14arxiv,wu2014low}.
A common cryptographic technique suitable for this purpose is scrambling (also known as random permutation), which has been widely used in the field of multimedia security \cite{dufaux2008scrambling,Wenjun:FrequencyScrambling:TM03}.
It should be noted that the works mentioned here and Sec.~\ref{subsec:C} are based on the RIP theory.
Here, we treat the measurement matrix as Gaussian matrix for simplicity\footnote{This simplification will not affect the security level of the discussed product cipher.}.}

{Roughly speaking, existing works utilizing scrambling for MTS usage of CS can be divided into two classes\footnote{Note that embedding scrambling in the time domain actually brings no benefit to security enhancement, but it helps the construction of a structural sampling ensemble \cite{Do:TSP:SRM12}.}:
\begin{enumerate}[I.]
\item Scrambling is performed on the measurements, such as \cite{zeng2012scrambling,huang2015compression};
\item Scrambling is done in the frequency domain, such as \cite{yushu:SCS:14arxiv,wu2014low}.
\end{enumerate}
The scrambling process can be characterized by a permutation matrix, which is a square binary matrix that has exactly one non-zero element with value $1$ in each row and each column and $0$s elsewhere. }

{
According to Eq.~(\ref{eq:samplingspace}), class I CS-based product cipher can be expressed as
\begin{IEEEeqnarray}{rCl}
\label{eq:encryptionBI}
\mathbf{\hat{y}} = \mathbf{P}_{K}\mathbf{y} = \mathbf{P}_{K}\mathbf{\Phi}_{K}\mathbf{x} =
\mathbf{P}_{K}\mathbf{\Phi}_{K} \mathbf{\Psi} \mathbf{s}  ,
\end{IEEEeqnarray}
where $\mathbf{x}$ is a $k$-sparse signal with dimension $M$ to be sampled (encrypted), $\mathbf{\Psi}$ is a orthnormal sparsifying basis, $\mathbf{P}_{K}$ is a $K \times K$ permutation matrix, $\mathbf{\Phi}_{K}$ is the Gaussian ensemble and $\mathbf{\hat{y}}$ is the ciphertext to be transmitted or store.
A difference between this class of product cipher and the basic CS-based ciphers
is that the (equivalent) secret key for the product cipher is the permutation matrix $\mathbf{P}_{K}$ and the measurement matrix $\mathbf{\Phi}_{K}$ while only measurement matrix can be utilized as the key in basic CS-based ciphers. Ideally (from the designer's point of view), the decoding (decryption) is composed of a two-step reconstruction, i.e.,
\begin{IEEEeqnarray}{rCl}
\mathbf{y} &=& \mathbf{P}_{K} \mathbf{\hat{y}}, \nonumber \\
  \min \|\mathbf{{s}}\|_1~~\text{subject to }\mathbf{y} &=&  \mathbf{\Phi}_{K} \mathbf{\Psi} \mathbf{s}. \nonumber
\end{IEEEeqnarray}
However, since both $\mathbf{P}_{K}$ and $\mathbf{\Psi}$ are orthonormal, $\mathbf{P}_{K}\mathbf{\Phi}_{K} \mathbf{\Psi}$, which is a rotation of $\mathbf{\Phi}_{K}$, possess the distribution of a Gaussian ensemble. Governed by the RIP theory, we can simplify the decoding as a single-step optimization
\begin{IEEEeqnarray}{rCl}
 \min \|\mathbf{{s}}\|_1~~\text{subject to }\mathbf{\hat{y}} &=& \mathbf{P}_{K}\mathbf{\Phi}_{K} \mathbf{\Psi} \mathbf{s}  = \mathbf{P}_{K}\mathbf{\Phi}_{K}\mathbf{x}. \nonumber
\end{IEEEeqnarray}
An unauthorized decoder, who can collect ciphertext for any plaintext in CPA scenario, submits a series of artificial signals
$\{\mathbf{x}_j \}_{j=1}^{M} = \{(0, \cdots, 0, 1_j, 0, \cdots, 0)^T\}_{j=1}^M$ to the encryption oracle and concludes
$\mathbf{P}_{K}\mathbf{\Phi}_{K} = \left[\mathbf{\hat{y}}_1, \cdots, \mathbf{\hat{y}}_M \right]$
using Eq.~(\ref{eq:encryptionBI}). It is clear that any further using of the same measurement and permutation matrices for security purpose is doomed to fail.
}

{
For the class II CS-based product ciphers, the same treatment can be applied. According to model (\ref{eq:samplingfrequency}), we can rewrite the encoding (encryption) process as
\begin{equation}
\mathbf{\hat{y}} = \mathbf{\Phi}_{K} \mathbf{P}_{K} \mathbf{s} = \mathbf{\Phi}_{K} \mathbf{P}_{K} \mathbf{\Psi}^{-1}\mathbf{x}. \nonumber
\end{equation}
Once again, $\mathbf{\Phi}_{K} \mathbf{P}_{K}$ can jointly working as the measurement matrix and
it can be revealed by $M$ independent chosen plaintexts and their corresponding ciphertexts.
}

{
In the following discussion, we will explain how scrambling (known as ``acceptable¡± permutation in  \cite{Fang:permutation:TSP13}) relaxes the RIP requirement of the measurement matrix for $2$D sparse signals.
Without loss of generality, let $\mathbf{X} = \left[\mathbf{X}_{i,j}\right]_{i=1, j=1}^{n, n}$ be a $2$D signal sparse in the canonic sparsifying basis and $\mathbf{k} = (k_1, k_2, \cdots, k_n)$ be a row vector whose entry denotes the number of nonzero elements of the columns of $\mathbf{X}$. A column by column sampling process of $\mathbf{X}$ can be summarized as
\begin{equation}
\mathbf{Y} = \left[\mathbf{y}_1, \mathbf{y}_2, \cdots, \mathbf{y}_n \right]= \mathbf{\Phi} \mathbf{X} = \mathbf{\Phi} \left[\mathbf{x}_1, \mathbf{x}_2, \cdots, \mathbf{x}_n \right], \nonumber
\end{equation}
or equivalently
\begin{IEEEeqnarray}{rCl}
\vec(\mathbf{Y}) = \left[\mathbf{y}_1, \mathbf{y}_2, \cdots, \mathbf{y}_n \right]^T=
\mathbf{\bar{\Phi}} \vec(\mathbf{X}) = \mathbf{\bar{\Phi}} \left[\mathbf{x}_1, \mathbf{x}_2, \cdots, \mathbf{x}_n \right]^T, \nonumber
\end{IEEEeqnarray}
where
\begin{IEEEeqnarray}{rCl}
\mathbf{\bar{\Phi}} = \left[
\begin{array}{cccc}
\mathbf{\Phi}   &                &               &\\
               & \mathbf{\Phi}   &               &\\
               &                &\ddots         &\\
               &                &               &\mathbf{\Phi}
\end{array} \nonumber \right].
\end{IEEEeqnarray}
The corresponding parallel (column by column) reconstruction is given by
\begin{IEEEeqnarray}{rCl}
\label{eq:PCSreconstruction}
 \min \|\mathbf{x}_j\|_1~~\text{subject to }\mathbf{y}_j &=&\mathbf{\Phi} \mathbf{x}_j,
\end{IEEEeqnarray}
where $j\in \{1, 2, \cdots, n\}$ and $\mathbf{\Phi}$ being a typical RIP measurement matrix with
$O(\|\mathbf{k}\|_{\infty} \cdot \log n )$ rows. As we can see, the accurate reconstruction is proportional to $\|\mathbf{k}\|_{\infty}$ \cite{Fang:permutation:TSP13}. The smaller $\|\mathbf{k}\|_{\infty}$ is, the fewer rows
$\mathbf{\Phi}$ require for correct recovery or the worse RIP constant $\mathbf{\Phi}$ can stand.}

{
The remaining work is to demonstrate that $\|\mathbf{k}\|_{\infty}$ of $\mathbf{X}$ will decrease with large probability if $\mathbf{X}$ is randomly scrambled.
Let $\vec(\mathbf{\bar{X}}) = P\cdot \vec(\mathbf{X})$ and $\mathbf{\bar{k}}=(\bar{k}_1, \cdots, \bar{k}_n)$ be the sparsity vector of $\mathbf{\bar{X}}$, we define an acceptable permutation as follows:
\begin{Definition}
A $n^2 \times n^2$ permutation $P$ is said to be acceptable if the following two rules are satisfied:
\begin{enumerate}
\item the expectations of the column sparsity of $\mathbf{\bar{X}}$ are the same, i.e., each column expects the same sparsity level;
\item the probability that $\|\mathbf{\bar{k}}\|_{\infty}$ deviates from the expected sparsity level observe a power law decay.
\end{enumerate}
\end{Definition}
The following property demonstrates the role of (secret) random scrambling for $2$D signals which is sparse in space. By swapping time and frequency, reconstruction model (\ref{eq:PCSreconstruction}) can be applied to natural $2$D signals, such as images.
The examples demonstrating this phenomenon will be provided in Sec.~\ref{sec:Sec4}.}
{
\begin{Property}
\label{pro:permutation}
Uniform random permutation is an \textit{acceptable} permutation for any $n\times n$ $2$D sparse signal $\mathbf{X}$.
\end{Property}}
\begin{IEEEproof}
{
To prove this, we recall that uniform random permutation refers to choosing a permutation from all the $(n^2)!$ candidates with equal probability.
In other words, each non-zero entry of $\mathbf{X}$ will appear at any location of $\mathbf{\bar{X}}$ with probability $1/n^2$ when $\mathbf{X}$ is processed by uniform random permutation.}

{
Since there are $\|\mathbf{k}\|_1$ non-zero entries of $\mathbf{X}$ in total, each entry of its permutated version is nonzero with probability $\|\mathbf{k}\|_1/n^2$. Apparently,
the expected sparsity level of $\mathbf{\bar{x}}_j$ is $n\times \frac{\|\mathbf{k}\|_1}{n^2}= \|\mathbf{k}\|_1/n$, which meets the requirements of rule~1).}

{
Treat each column of $\mathbf{\bar{X}}$ as realization of $n$ independent, identically distributed random variables,
the probability that $\|\mathbf{\bar{k}}\|_{\infty}$ deviates from the expectation $\|\mathbf{k}\|_1/n$  by $t$ can be characterized by
\vspace{-0.25cm}
\begin{IEEEeqnarray}{rCl}
\IEEEeqnarraymulticol{3}{l}{
Prob((\|\mathbf{\bar{k}}\|_{\infty} - \|\mathbf{k}\|_1/n \geq t)
}\nonumber\\
&=    &  Prob( (\max_j (\bar{k}_j) -\|\mathbf{k}\|_1/n ) \geq t) \nonumber\\
&\leq &  Prob( ( \bar{k}_j - \|\mathbf{k}\|_1/n) \geq t) \nonumber\\
&\leq &  e^{-{2nt^2}},\nonumber
\end{IEEEeqnarray}
where the last inequality is obtained by applying Hoeffding inequality. 
Hence finishes the proof.
}
\end{IEEEproof}

 \subsection{{Concatenation of CS and DRPE}}
\label{subsec:C}
{
As one of the optical information processing technique, image encryption using DRPE has received a lot of research attention since its first appearance in \cite{Refregier:DRPE:OL95,javidi1999method}. This cipher was found insecure against various plaintext attacks \cite{carnicer2005vulnerability,frauel2007resistance}.
In a different context, CS offers a new approach for hologram compression and sensing in the optical domain \cite{clemente2013compressive,rivenson2010compressive}. On the one hand, the concatenation of CS and DRPE enjoys a all-optical implementation and substantially data volume reduction. On the other hand, the secrecy provided by CS may enhance the security level of DRPE, and vice visa.
These reasons making cascading CS and DRPE a noticeable alternative to support the MTS usage of CS. In the following discussion, we will point out that the later argument is questionable in MTS scenario since the CPA complexity of this model is exactly the same as that of the basic CS model.
}

{
Considering a discrete and bounded\footnote{This always holds true given that continuous data can be adequately sampled.} $2$D data $\mathbf{I}=[\mathbf{I}_{i,j}]$, the DRPE encryption can be formulated as
\begin{IEEEeqnarray}{rCl}
\mathbf{C}_{i,j} &=&
\mathcal{IF}\left( \mathcal{FT} \left( \mathbf{I}_{i,j} \cdot \exp(j2\pi p_{i,j}) \right) \cdot \exp(j2\pi q_{u,v}) \right), \nonumber
\end{IEEEeqnarray}
where the random spatial phase mask $\mathbf{P}=[\exp(j2\pi p_{i,j})]$ and the random frequency phase mask $\mathbf{Q}=[\exp(j2\pi q_{u,v})]$ are the secret keys, and $\mathcal{FT}(\mathbf{X}) =\mathbf{F}\mathbf{X} \mathbf{F}^*$ with $\mathbf{\cdot}^*$ being the conjugate transpose and $\mathcal{IF}$ being the inverse Fourier transform. The DRPE decryption is omitted here since it is similar to the encryption process. With these notations, we can also divide the encryption schemes based on concatenation of CS and DRPE into two classes:
\begin{enumerate}[I.]
\item CS encryption followed by DRPE \cite{deepan2014multiple};
\item DRPE followed by CS encryption \cite{rawat2015compressive,li2015compressive}.
\end{enumerate}
}

{
Considering a $2$D image $\mathbf{X}$ with $M = n\times n $ pixels is sensed by CS with $K=m \times m$ measurements, the algorithms of class I can be modeled as a separate two-step process, i.e.,
\begin{IEEEeqnarray}{rCl}
\vec(\mathbf{Y}) = \mathbf{\Phi} \vec(\mathbf{X}), \nonumber \\
\mathbf{C} = \mathcal{IF}\left( \mathcal{FT} \left( \mathbf{Y}_{i,j} \cdot \exp(j2\pi p_{i,j}) \right) \cdot \exp(j2\pi q_{u,v}) \right) ,
\label{eq:DRPE}
\end{IEEEeqnarray}
where $\mathbf{\Phi}_{m^2 \times n^2}$, $\mathbf{P}_{m \times m}=[\exp(j2\pi p_{i,j})]$ and $\mathbf{Q}_{m \times m}=[\exp(j2\pi q_{u,v})]$ serve as the (equivalent) secret key in the whole process and $\mathbf{C}$ is the ciphertext to deliver or display. As claimed in \cite{deepan2014multiple}, decoding $\mathbf{C}$ 
should observe a separate DRPE decryption and CS reconstruction, or by a reversed order in algorithms belonging to class II \cite{rawat2015compressive,li2015compressive}.
 As such, it is demonstrated that an unauthorized user
who cannot access full knowledge of $\mathbf{\Phi}$, $\mathbf{P}$ and $\mathbf{Q}$ is not able decrypt $\mathbf{X}$ \cite{deepan2014multiple,rawat2015compressive,li2015compressive}.
}

{
We investigate the real strength against  CPA for the approaches mentioned above by first rewriting Eq.~(\ref{eq:DRPE}) as a matrix form \cite{frauel2007resistance}, i.e.,
\begin{IEEEeqnarray}{rCl}
\vec(\mathbf{C}) & = & \mathbf{T} \vec(\mathbf{Y}), \nonumber \\
                 & = & \mathbf{\bar{F}}^* \mathbf{\bar{Q}} \mathbf{\bar{F}} \mathbf{\bar{P}} \cdot \vec(\mathbf{Y}), \nonumber
\end{IEEEeqnarray}
where $\mathbf{\bar{F}}_{m^2\times m^2}$ is the Kronecker product of the Fourier matrices $\mathbf{F}^*$ and $\mathbf{F}$, $\mathbf{\bar{P}}_{m^2\times m^2} =\diag(\vec(\mathbf{P}))$ and
$\mathbf{\bar{Q}}_{m^2\times m^2} =\diag(\vec(\mathbf{Q}))$ are the DRPE secret key. By construction, $\mathbf{\bar{P}}$ and $\mathbf{\bar{Q}}$ are unitary matrices. So, it is concluded $\mathbf{T}$ is also a unitary matrices.
In this concern, $\mathbf{T}\mathbf{\Phi}$ must be a RIP matrix and thus a single-step optimization can be formulated as\footnote{We note that the multiple measurement vector CS model \cite{duarte2005distributed}  should be adopted since $\mathbf{T}$ is a complex matrix.}
\begin{IEEEeqnarray}{rCl}
 \min \|\mathbf{\Psi}^{-1}\cdot \vec(\mathbf{X})\|_1~~
 \text{subject to }
 \vec(\mathbf{C}) &=& \mathbf{T}\mathbf{\Phi} \vec(\mathbf{X})  . \nonumber
\end{IEEEeqnarray}
Once again, the attacker who works under CPA assumption can retrieve $\mathbf{T}\mathbf{\Phi}$ faithfully from $M$ independent plaintexts and the corresponding ciphertexts. Moreover, he can use this information to decode (decrypt) any subsequent ciphertexts.
Similarly, we can apply the analyses to class II algorithms and obtain the same conclusion.
}

\section{The Proposed Scheme}
\label{sec:secIII}
{
As reviewed in the previous section, existing proposals \cite{zeng2012scrambling,huang2015compression,yushu:SCS:14arxiv,wu2014low,deepan2014multiple,rawat2015compressive,li2015compressive} targeting the MTS usage of CS as joint sampling and data protection mechanism fail to resist plaintext attacks. Similarly, it can be concludes that cascading CS, scrambling and DRPE also suffer from the same defect, such as the one suggested in \cite{liu2013optical}. The underlying reason is that all these three cryptographic primitives are linear and we can always translate the encoding components to a (equivalent) RIP-based measurement matrix.
Therefore, the key question is whether it is possible to construct a more secure CS-based product cipher without introducing any computing-intensive cryptographic primitives. We will give a positive solution to this problem by switching from the RIP measurement matrix construction to the RIPless matrix construction. We start with the following example.}

\begin{figure}[!htb]
\centering
\begin{minipage}[t]{\figwidthb}
\centering
\includegraphics[width=\figwidthb]{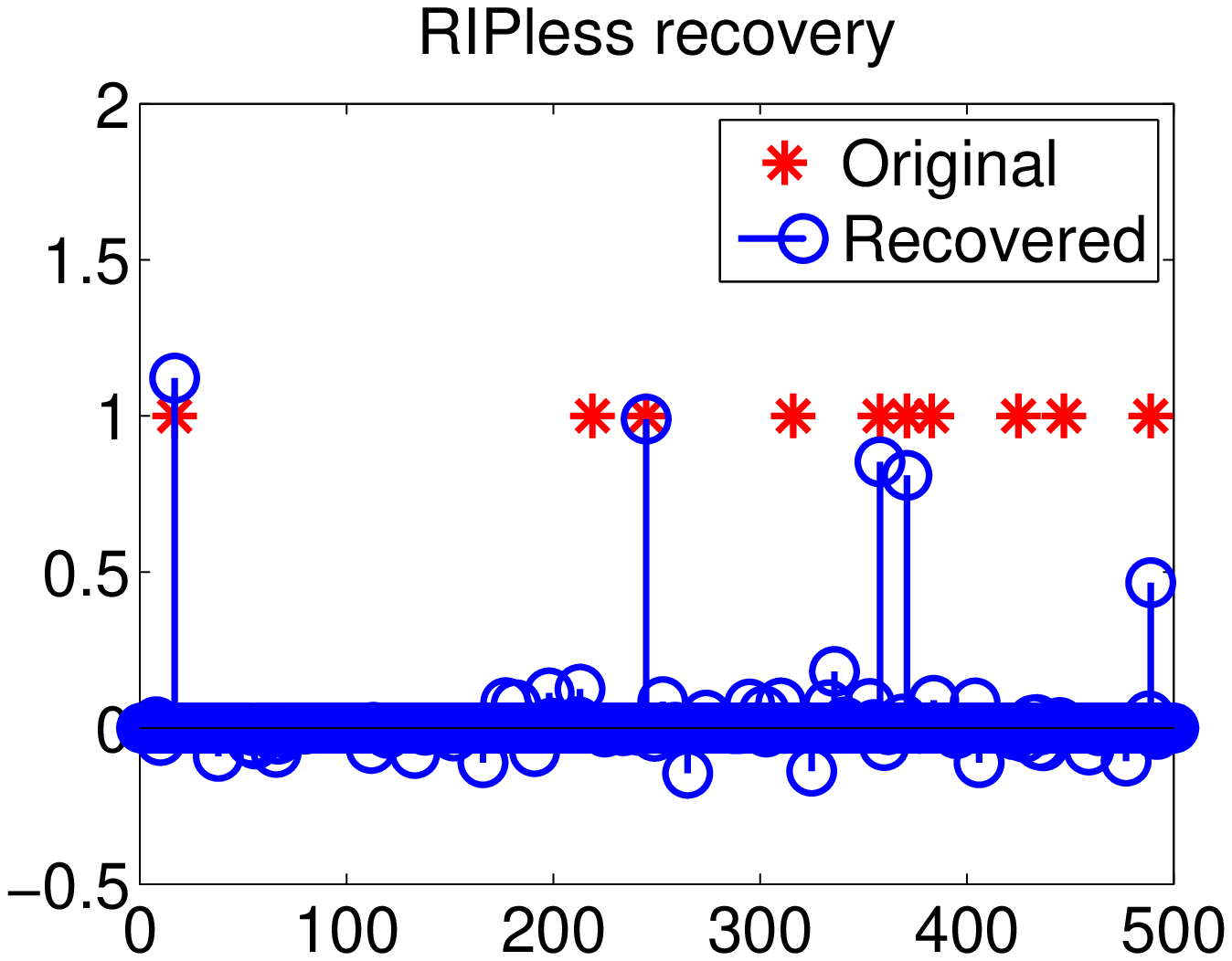}
\end{minipage}
\centering
\begin{minipage}[t]{\figwidthb}
\centering
\includegraphics[width=\figwidthb]{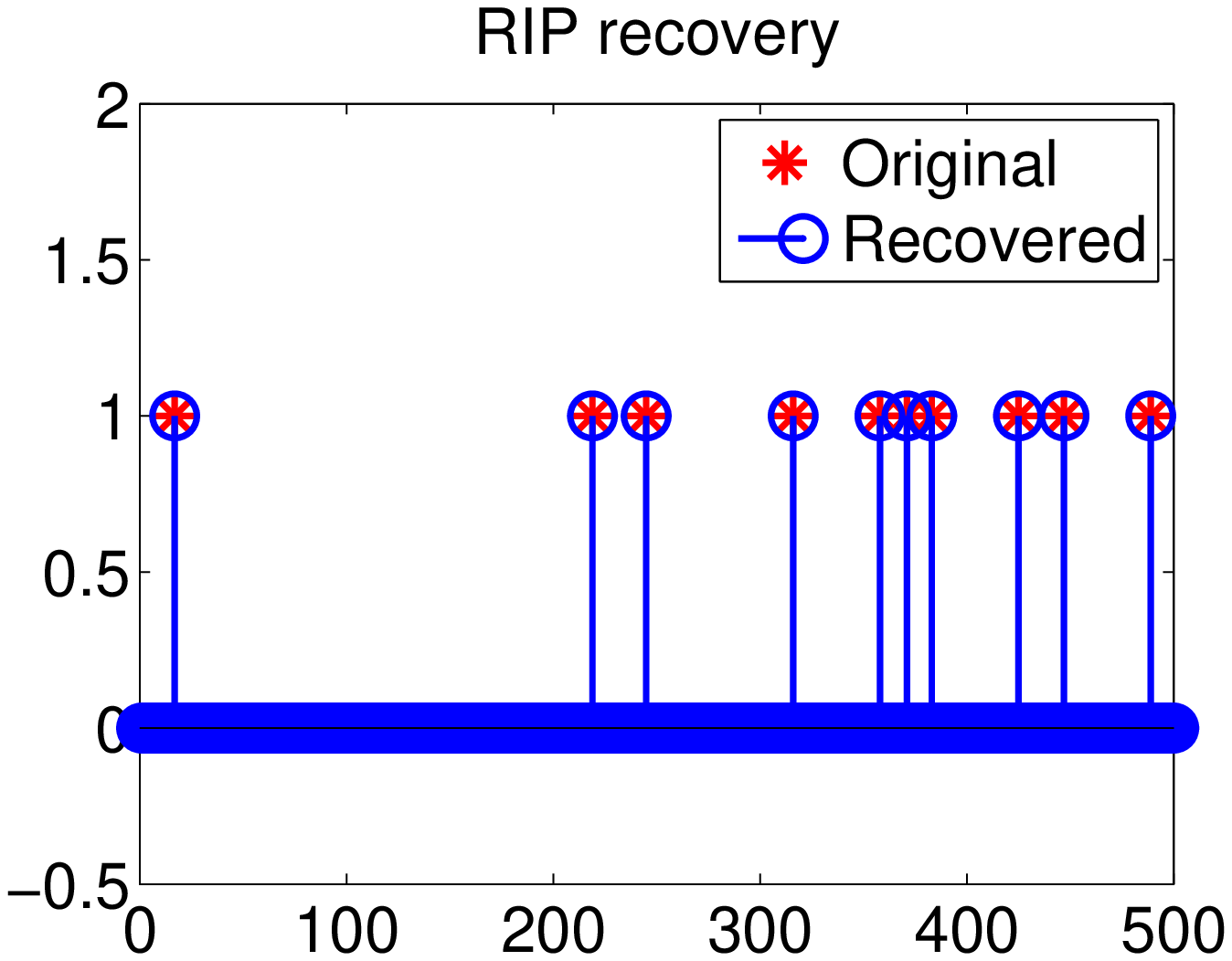}
\end{minipage}
\caption{
Example of RIPless reconstruction and RIP reconstructions.}
\label{fig1:RIPandRIPless}
\end{figure}

{
Consider a column vector $\mathbf{x}$ of length $M=500$ taking values from $\{0, 1\}$ has a sparsity level $k=10$. Let $F$ denote an independent multivariate antipodal distribution, which is given by $F = \{\pm d_1\} \times \{\pm d_2\} \times \cdots \times \{\pm d_M\}$ with $Prob(d_j)=Prob(-d_j)=1/2$ and $\{d_j\}_{j=1}^{M}$ be positive integers. We take $60$ sensing vectors\footnote{Here, we take $K=60$ because $K>4k$ is an empirical threshold for exact CS recovery in the RIP theory \cite{Candes:IntroCS:SPM08}.}
from this distribution and get a measurement matrix $\mathbf{\Phi}$ which is further used to sample $\mathbf{x}$.
By Definition~1, $\mathbf{\Phi}$ cannot guarantee energy-preserving property thus it is a non-RIP matrix.
By construction, we have $\theta = O(\max_j(d_j)/\min_j(d_j))$ and
\begin{IEEEeqnarray}{rCl}
\mu(F) &\geq& \max_{1\leq i \leq M } |<\mathbf{\phi}^T, \mathbf{e}_i>| \nonumber \\
        &=& max_j(d_j) \nonumber.
\end{IEEEeqnarray}
In summary, $\mu(F) \theta=O(\max_j(d^2_j)/\min_j(d_j))$ is a non-negligible term and the following straightforward recovery
dominated by RIPless theory (see Theorem~1 for detail)
\begin{equation}
  \min \|\mathbf{\bar{x}}\|_1~~\text{subject to }  \mathbf{y}=\mathbf{\Phi}  \mathbf{\bar{x}} \nonumber
\end{equation}
returns a solution $\mathbf{\bar{x}}\neq \mathbf{x}$. Set $\mathbf{A} = \mathbf{\Phi}\mathbf{D} = \mathbf{\Phi}\cdot \diag(1/d_1,\cdots, 1/d_M)$, the reconstruction can also transformed to a two-step reconstruction compliance with RIP theory after realizing that $\mathbf{A}$ is a Bernoulli matrix, i.e.,
\begin{IEEEeqnarray}{rCl}
  \min \|\mathbf{\hat{x}}\|_1~~\text{subject to }\mathbf{y} &=& (\mathbf{A}\mathbf{D}^{-1}) \mathbf{x}= \mathbf{A} \mathbf{\hat{x}}, \nonumber \\
        \mathbf{\bar{x}} &=& \mathbf{D} \mathbf{\hat{x}}. \nonumber
\end{IEEEeqnarray}
We compare the recovery techniques described above. Figure~\ref{fig1:RIPandRIPless} depicts a typical reconstruction result with $d_j \in [1, 60]$, from which we can see that the recovery in the RIP case is exact but the RIPless case is not due to a lack of sufficient measurements.
}


{
The above example provides a preparatory understanding of how a RIPless matrix construction can be transformed to a RIP one. Still, it cannot be considered as a good CS-based cipher since an attacker can reveal $\mathbf{D}$ from $\mathbf{\Phi}$ by $d_j = |\mathbf{\Phi}_{i,j}|$. Moreover, this technique only works for vector who is sparse in the canonical basis, which is not practical for real signals.
In this concern, we apply this finding to the CS model~(\ref{eq:samplingfrequency}) and devise a so called bi-level protected CS model in a way that the measurement matrix is non-RIP and the reconstruction works under RIP theory. }

{The BLP-CS model will be described in}
Sec.~\ref{subsec:newmodel}, which can be viewed as product of the CS-based cipher and a transform encryption. Then we propose two methods for key-related sparsifying transformation design, namely, \textit{Type I Secret Basis} and \textit{Type II Secret Basis}.


\subsection{Bi-level Protection Model}
\label{subsec:newmodel}
\begin{figure}[!htb]
\centering
\begin{minipage}[t]{\imagewidtha}
\centering
\includegraphics[width=\imagewidtha]{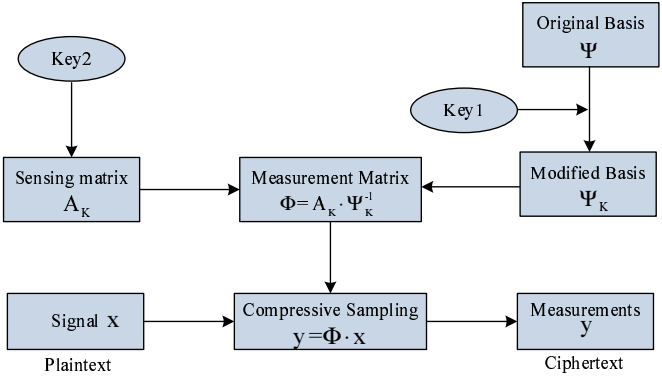}
\end{minipage}
\caption{Block diagram of BLP-CS.}
\label{fig:newmodel}
\end{figure}

 {The} block diagram of this model is shown in Fig.~\ref{fig:newmodel}, where we suggest using key-dependent sensing matrix, $\mathbf{A}_K$, and secret-related sparsifying basis, $\mathbf{\Psi}_K$, to determine the measurement matrix $\mathbf{\Phi}= \mathbf{A}_K\mathbf{\Psi}^{-1}_K$. 
{Recalling the above example}, we are interested in {the phenomenon that the measurement matrix $\mathbf{\Phi}$ does not satisfy the RIP requirement, while the key-dependent sensing matrix $\mathbf{A}_K$ itself is a RIP matrix.}
{Referring to Eq.~(\ref{eq:samplingfrequency})}, the sampling procedure can be expressed as
\begin{equation*}
\mathbf{y} = \mathbf{\Phi}\mathbf{x}
=  \mathbf{A}_K\mathbf{\Psi}^{-1}_K (\mathbf{\Psi}_K\mathbf{s}) = \mathbf{A}_K \mathbf{s}.
\end{equation*}
{It should be noted that the number of measurements (sampling rate) is on the order of $(k \log M)$ even though $\mathbf{\Phi}$ is a non-RIP measurement matrix. This number of measurements fails to meet the minimum  requirement defined in Theorem~\ref{theorem:RIPless}, thus makes the correct decoding from $\mathbf{\Phi}$ an impossible task.
}

{To correctly decode (decrypt) $\mathbf{y}$,} a legitimate user should first derive $\mathbf{A}_K$ and $\mathbf{\Psi}_K$ from the key scheduling process and then {refer to the following two-step reconstruction}
\begin{IEEEeqnarray}{rCl}
  \min \|\mathbf{{s}}\|_1~~\text{subject to }\mathbf{y} &=&  \mathbf{\Phi}  \mathbf{x}=\mathbf{A}_K  \mathbf{{s}}, \nonumber \\
  \mathbf{x} &=& \mathbf{\Psi}_K \mathbf{s}. \nonumber
\end{IEEEeqnarray}
or equivalently
\begin{IEEEeqnarray}{rCl}
  \min \|\mathbf{\Psi}_K^{-1}\mathbf{x}\|_1~~\text{subject to }\mathbf{y} &=&  \mathbf{\Phi}  \mathbf{x}, \nonumber
\end{IEEEeqnarray}
{To fulfill the security requirement, the remaining task is to design two matrices $\mathbf{A}_K$ and $\mathbf{\Psi}_K$ satisfying:
\begin{enumerate}[RULE a.]
\item $\mathbf{A}_K$ is a key-related matrix satisfy RIP;
\item $\mathbf{\Psi}_K$ is a key-related sparsifying basis;
\item $\mathbf{A}_K\mathbf{\Psi}^{-1}_K$ is a structural non-RIP matrix.
\end{enumerate}
}
The work of designing a RIP matrix is trivial since it is already clear that Guussian/Bernoulli \cite{Candes:CS:TIT06} and structurally random matrices \cite{Do:TSP:SRM12} are competent for this task with overwhelming probability. Therefor, we focus our attention on the designing of $\mathbf{\Psi}_K$ in the following discussions. {It is worth mentioning that the work of designing $\mathbf{\Psi}_K$ satisfying RULE~b (also known as transform encryption) is very popular in the filed of multimedia encryption, examples can be found in \cite{zeng2014perceptual,pande2012secure,pande2013securing}. However, the work of designing $\mathbf{A}_K$ and $\mathbf{\Psi}_K$ satisfying RULE~c is totally new.}

\subsection{Type I Secret Basis}
\label{sec:TypeI}
{
The first type of {secret} basis that drawn our attention is the parameterized construction of some familiar transform, such as parameterized discrete wavelet transform (DWT) \cite{engel2005parameterized,pande2012secure} and directional discrete cosine transfrom (DCT) \cite{yeung2012new,zeng2014perceptual}. Here, we present a parameterized transform based on Fractional Fourier Transform (FrFT) as an example. 
}

The use of FrFT for security purpose can be dated back to year $2000$, when
Unnikrishnan \textit{et al.} \cite{Joseph:Frft:OL00} suggested to use FrFT for DRPE instead of the ordinary Fourier transform \cite{Refregier:DRPE:OL95},
in order to benefit from its extra degrees of freedom provided by the fractional orders.
Generally speaking, performing an order $\alpha$ FrFT on a signal can be viewed as a rotation operation on the time-frequency or space-frequency distribution at an angle $\alpha$.
Though FrFT is very popular in optics for its easy implementation, it is not preferred in digital world since complex numbers always cause extra computational load.

To this end, Venturini \textit{et al.} proposed a method to construct Reality-Preserving FrFT of arbitrary order \cite{Venturini:RealFrct:04}.
Here, we deduce the Reality-Preserving Fractional Cosine Transform (RPFrCT) by the virtue of their method. Denote the discrete cosine transform \cite{Gianfranco:Ftct:TSP02}
of size $n\times n$ by
\begin{equation*}
\mathbf{C} =\left( \frac{1}{\sqrt{n}} \epsilon_l \cos(2 \pi \frac{(2i+1)l}{4n})\right),
\end{equation*}
where $i = 0\sim n-1$, $l= 0\sim n-1$, $\epsilon_0 =1$ and $\epsilon_l= \sqrt{2}$ for $l>0$.
The unitary property of $\mathbf{C}$ assures that it can be diagonalized as
\begin{equation}
\label{eq:Dct}
\mathbf{C} = \mathbf{U} \mathbf{\Lambda} \mathbf{U}^*,
\end{equation}
where 
$\mathbf{U} = \{\mathbf{u}_i\}_{i=1}^n$ is composed of $n$ orthonormal eigenvectors, i.e.,
$\mathbf{u}^*_m \mathbf{u}_i = \delta_{mi}$ and $\mathbf{\Lambda} = \diag(\lambda_1, \cdots, \lambda_i,\cdots , \lambda_n)$ with $\lambda_i = \exp(j \varphi_i)$.
Replace $\lambda_i$ with its $\alpha$-th power $\lambda_i^\alpha$ in Eq.~(\ref{eq:Dct}), we can express the Discrete Fractional Cosine Transform (DFrCT) matrix
$\mathbf{C}_\alpha$ of order $\alpha$ in the compact form
\begin{equation*}
\mathbf{C}_\alpha =\mathbf{U} \mathbf{\Lambda}^\alpha \mathbf{U}^*.
\end{equation*}
Having defined $\mathbf{C}_\alpha$, we can derive the RPFrCT matrix $\mathbf{R}_\alpha$ as follows:
\begin{itemize}

\item For any real signal $\mathbf{x}=\{x_l\}_{l=1}^{M}$ of length $M$ ($M$ is even), construct a complex signal of length $M/2$ by
    \begin{equation*}
        \mathbf{\widetilde{x}} = \{x_1+ jx_{M/2+1}, x_2+ jx_{M/2+2}, \cdots, x_{M/2}+ jx_{M} \}.
    \end{equation*}
\item Compute $\mathbf{\widetilde{y}} = \mathbf{B}_\alpha \mathbf{\widetilde{x}}$, where $\mathbf{B}_\alpha$ is a DFrCT matrix of size $(M/2\times M/2)$, namely,
    $\mathbf{B}_\alpha =  \mathbf{C}_{\alpha, M/2}$.
\item Determine the RPFrCT matrix $\mathbf{R}_\alpha$ by
    \begin{eqnarray}
    \mathbf{y} & = & (\Re(\mathbf{\widetilde{y}}), \Im(\mathbf{\widetilde{y}}) )^T \nonumber\\
               & = & \left(
                    \begin{array}{c}
                        \Re(\mathbf{B}_\alpha) \Re(\mathbf{\widetilde{x}}) - \Im(\mathbf{B}_\alpha) \Im(\mathbf{\widetilde{x}})  \nonumber\\
                        \Im(\mathbf{B}_\alpha) \Re(\mathbf{\widetilde{x}}) + \Re(\mathbf{B}_\alpha) \Im(\mathbf{\widetilde{x}})   \nonumber\\
                        \end{array}
                   \right)\\
                & = & \left(
                    \begin{array}{cc}
                        \Re(\mathbf{B}_\alpha) & -\Im(\mathbf{B}_\alpha)  \nonumber\\
                        \Im(\mathbf{B}_\alpha) & \Re(\mathbf{B}_\alpha)  \nonumber\\
                        \end{array}
                   \right) \cdot
                   \left(
                   \begin{array}{c}
                         \Re(\mathbf{\widetilde{x}}) \nonumber\\
                         \Im(\mathbf{\widetilde{x}}) \nonumber\\
                        \end{array}
                   \right)  \\
                & = &    \mathbf{R}_\alpha \mathbf{x} \nonumber.
    \end{eqnarray}
\end{itemize}

\begin{figure}[!htb]
\centering
\begin{minipage}[t]{\imagewidth}
\includegraphics[width=\imagewidth]{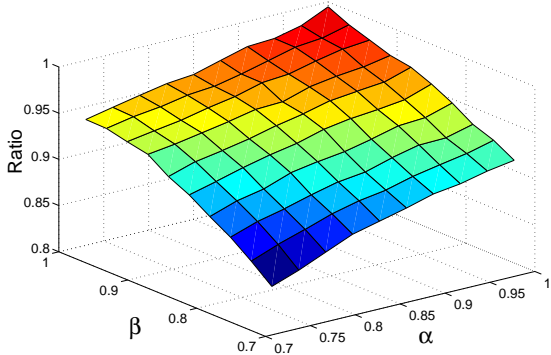}
\end{minipage}
\caption{Comparison between the recovery result of RPFrCT and DCT2 using the best $s$-term approximation at different fractional orders.}
\label{fig:FrCTDCT}
\end{figure}

From the construction process listed above, we can conclude that $\mathbf{R}_\alpha$ is orthogonal, reality preserving and periodic. 
Then, the Reality-Preserving Fractional Cosine Transform of a digital image $\mathbf{X}$
is given by
\begin{IEEEeqnarray}{rCl}
\label{eq:2dbasis}
\mathbf{S} = \mathbf{R}_\alpha \mathbf{X} \mathbf{R}^{T}_\beta,
\end{IEEEeqnarray}
where {$(\cdot)^T$ represents the transpose operator, }
$\alpha$ and $\beta$ are the orders of the Fractional Cosine Transform along $x$ and $y$ directions, respectively. {Equivalently, we can express this formula as
\begin{IEEEeqnarray}{rCl}
\vec(\mathbf{S}) = \mathbf{\Psi}^{-1} \vec(\mathbf{X}), \nonumber
\end{IEEEeqnarray}
where $\mathbf{\Psi}^{-1} = \mathbf{\Psi}^{T}=(\mathbf{R}_\beta \otimes \mathbf{R}_\alpha)$.
To study the sparsifying capability of the proposed parameterized basis,
}
we carried out experiments {on digital images} at different fractional orders $\alpha$ and $\beta$
by using the best $s$-term approximation, i.e., keep the $s$ largest coefficients and set the remaining ones to zero.
The recovered result of RPFrCT is compared with that of DCT$2$
using the ratio between their peak signal-to-noise ratios (PSNRs).
As expected, the sparsifying capability of RPFrCT raises when $\alpha$ or $\beta$ increases, as shown in Fig~\ref{fig:FrCTDCT}.
When $\alpha, \beta \in (0.9, 1]$, the sparsifying capability of RPFrCT is comparable to that of DCT2. {It is worth mentioning that a similar sparsifying capability was also observed when this transform is applied to $1$D signals \cite{Venturini:RealFrct:04}.}

\subsection{Type II Secret Basis}
\label{sec:TypeII}
{
 We have demonstrated a technique for parameterized sparsifying basis construction, where the free parameter can be used as the secret key in the BLP-CS model. In this way, the resultant basis satisfies RULE~b. However, it still suffers from the same CPA shown in Sec.~\ref{sec:II} since it fails to meet RULE~c. In the subsequent discussions, we propose three kind of operations on an existing basis to make it fulfill RULE c.
}
We start the deviation by defining equivalent sparsifying bases.
\begin{Definition}
Two basis matrices, $\mathbf{\Psi}$ and $\mathbf{\Psi}'$ are equivalent sparsifying bases if
$\mathbf{x} = \mathbf{\Psi}\mathbf{s} = \mathbf{\Psi}'\mathbf{s}'$, $\|\mathbf{s}\|_0 = \|\mathbf{s}'\|_0=k$ holds for any signal $\mathbf{x}$.
\end{Definition}

\begin{Property}
\label{pro:basischange1}
$\mathbf{\Psi}'$ and $\mathbf{\Psi}$ are equivalent sparsifying bases if
\begin{IEEEeqnarray}{rCl}
\mathbf{\Psi}' & = & \mathbb{F}_1(\mathbf{\Psi}) \nonumber\\
                & = &(d_1\mathbf{\psi}_1, d_2\mathbf{\psi}_2, \cdots, d_j\mathbf{\psi}_j, \cdots, d_M\mathbf{\psi}_M), \nonumber
\end{IEEEeqnarray}
where $\{d_j\}_{j=1}^M$
are non-zero constants and $\psi_j$ is the $j$-th column of $\mathbf{\Psi}$.
\end{Property}
\begin{proof}
Set $s'_j = \frac{1}{d_j} s_j$ and we have $\|\mathbf{s}\|_0 = \|\mathbf{s}'\|_0$.
\end{proof}
{We demonstrate that we are able to construct a non-RIP measurement matrix satisfying RULE c. Assume $\mathbf{\Psi}$ is an orthonormal basis and set
\begin{IEEEeqnarray}{rCl}
\mathbf{\Psi}' = \mathbf{\Psi}\mathbf{D} , \nonumber
\end{IEEEeqnarray}
where $\mathbf{D}=\diag(1/d_1, 1/d_2, \cdots, 1/d_M)$ and $\{d_j\}_{j=1}^{M}$
are positive integers drawn from certain distribution independently.
Let $\mathbf{A}$ denote a Gaussian matrix with i.i.d. entries and calculate $\mathbf{\Phi}$ as
\begin{IEEEeqnarray}{rCl}
\mathbf{\Phi} &=& \mathbf{A} (\mathbf{\Psi}\mathbf{D})^{-1} , \nonumber \\
                &=& \mathbf{A} \mathbf{D}^{-1} \mathbf{\Psi}^{T}. \nonumber
\end{IEEEeqnarray}
Once again, the effect of $\mathbf{\Psi}^{T}$ can be viewed as a rotation of $\mathbf{A} \mathbf{D}^{-1}$ in a $M$ dimensional space, which is energy preserving. By construction, $\mathbf{\Phi}$ is a non-RIP matrix.
}

\begin{Property}
\label{pro:basischange2}
$\mathbf{\Psi}'$ and $\mathbf{\Psi}$ are equivalent sparsifying bases if
\vspace*{-0.6\baselineskip}
\begin{equation*}
\mathbf{\Psi}' = \mathbb{F}_2(\mathbf{\Psi}) =  \mathbf{\Psi} \mathbf{P},
\end{equation*}
where $\mathbf{P}$ is a random permutation matrix.
\end{Property}
\begin{proof}
Since $\mathbf{\Psi}\mathbf{s} = \mathbf{\Psi}(\mathbf{P}\mathbf{P}^T)\mathbf{s} = \mathbf{\Psi}' (\mathbf{P}^T\mathbf{s}) =\mathbf{\Psi}'\mathbf{s}'$
, $\|\mathbf{s}'\|_0 = \|\mathbf{P}^T\mathbf{s}\|_0=\| \mathbf{s}\|_0$.
\end{proof}
{In the $1$D case, this property implies that random scrambling does not cause any loss of the sparsity level of any given signal. In the $2$D case, as we have shown in Sec.~\ref{subsec:B}, it helps to uniform the column (or row) sparsity level and thus flavors a parallel CS reconstruction technique,
which will be exemplified in Sec~\ref{sec:Sec4}.}

\begin{figure}[!htb]
\centering
\begin{minipage}[t]{\figwidth}
\centering
\includegraphics[width=\figwidth]{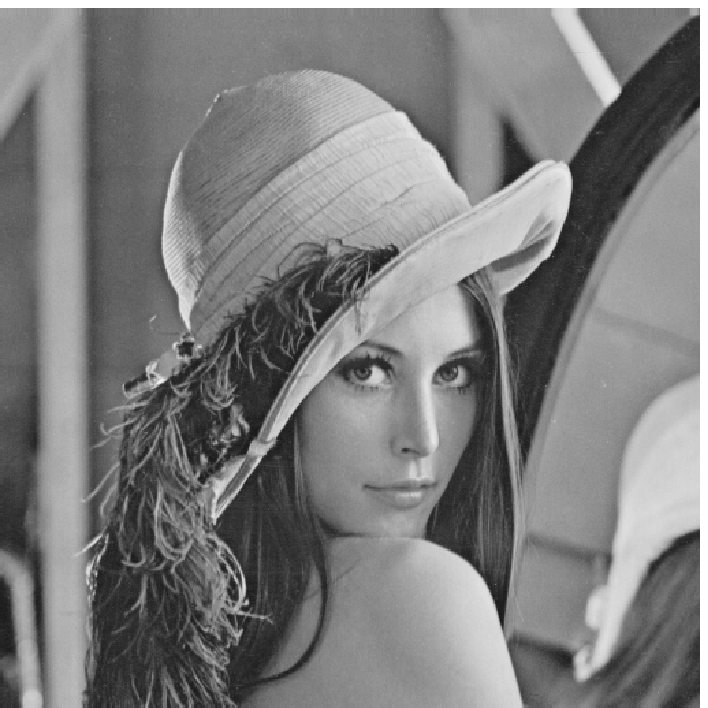}
a)
\end{minipage}\\
\centering
\begin{minipage}[t]{\figwidtha}
\centering
\includegraphics[width=\figwidtha]{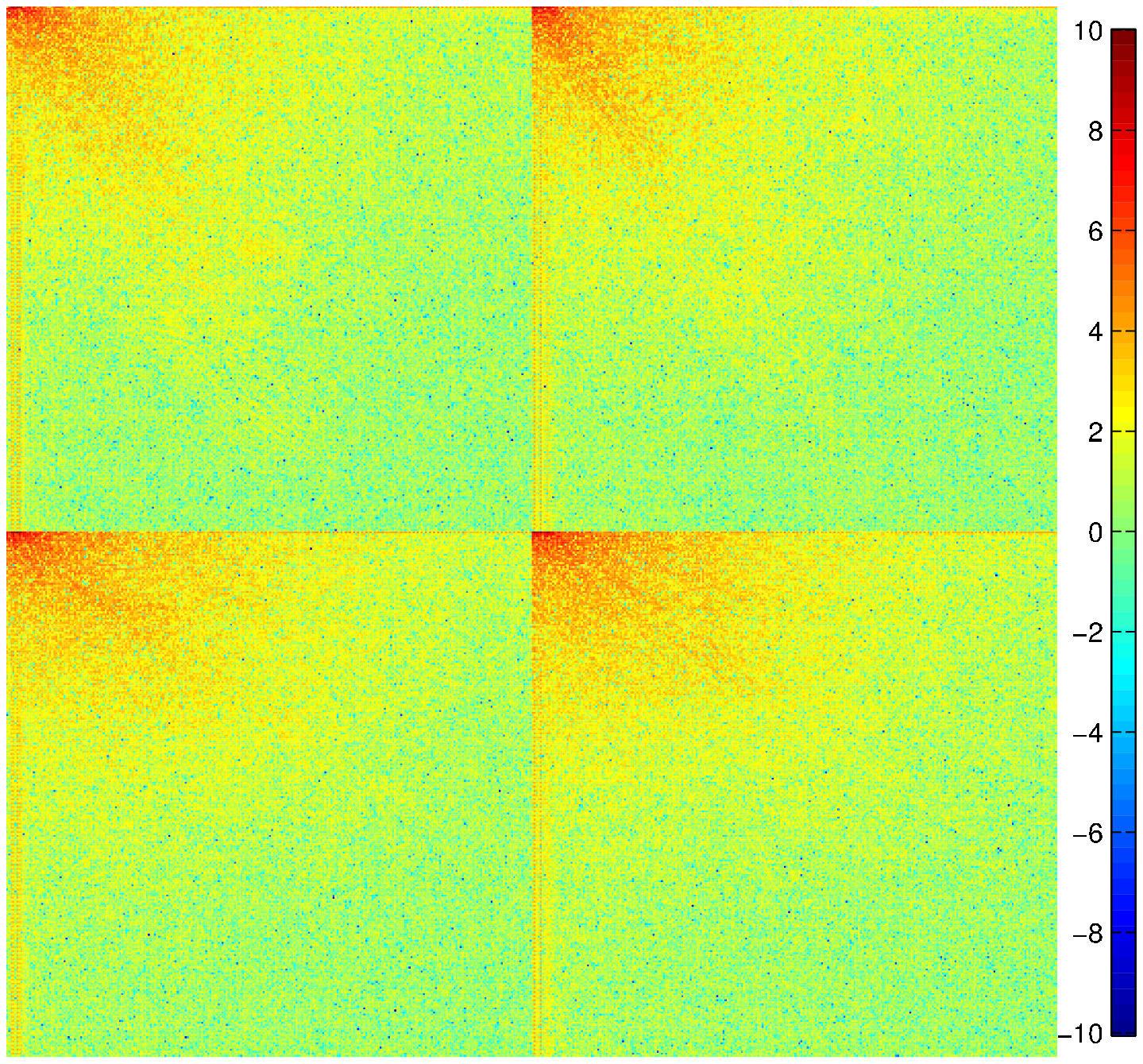}
b)
\end{minipage}
\caption{a)~Original image ``Lena"; b)~Energy distribution of RPFrCT coefficients of ``Lena" using logarithm base.}
\label{fig3:DCT2Energy}
\end{figure}

In addition, if we know or partially know that $\supp(\mathbf{s})$ is localized in a certain $k$-dimensional subspace
rather than uniformly distributed in $\mathbb{R}^N$, we can embed more secrets into the sparsifying basis, as stated in
Property~\ref{pro:basischange3}. Here we assume that $\mathbf{\Psi}$ is an orthonormal sparsifying basis for simplicity.
\begin{Property}
\label{pro:basischange3}
$\mathbf{\Psi}'$ and $\mathbf{\Psi}$ are equivalent sparsifying bases if
\vspace*{-0.6\baselineskip}
\begin{IEEEeqnarray}{rCl}
\mathbf{\Psi}' & = & \mathbb{F}_3(\mathbf{\Psi}) \nonumber \\
               & = & (\psi_1, \cdots, \psi_{j-1}, a\psi_j+b\psi_k,\psi_{j+1}, \cdots, \psi_M),   \nonumber
\end{IEEEeqnarray}
where $a,b$ are non-zero constants and $j, k \in \supp(\mathbf{s})$ or $j, k \notin \supp(\mathbf{s})$.
\end{Property}
\begin{proof}
Since $\mathbf{\Psi}$ is orthonormal, $s_j = (\mathbf{\psi}_j, \mathbf{x}) =\mathbf{\psi}_j^T \mathbf{x}$ and we know
$s_j = 0$ when $j \notin \supp(\mathbf{s})$. Then the proof for $j, k \notin \supp(\mathbf{s})$ is trivial. For $j, k \in \supp(\mathbf{s})$, set
$\mathbf{s}' =  (s'_1, s'_2, \cdots,s'_j, \cdots, s'_k, \cdots, s'_M)^T  \nonumber $
with
\vspace*{-0.6\baselineskip}
\begin{equation}
\label{eq:suppunchanged}
s'_i = \left\{
\begin{array}{rl}
s_i/a & \text{if } i = j,  \\
s_i - s_j{b}/{a} & \text{if } i = k, \\
s_i & \text{otherwise}.
\end{array} \right.
\end{equation}
Then we have
\vspace*{-0.6\baselineskip}
\begin{IEEEeqnarray*}{rCl}
\mathbf{x} & = & \mathbf{\Psi}\mathbf{s}  \\
& = & \sum_{\substack{i=1 \\ i\neq j,k}}  ^{N} s_i\mathbf{\psi}_i  + s_j \mathbf{\psi}_j +s_k \mathbf{\psi}_k \\
& = & \sum_{\substack{i=1 \\ i\neq j,k}}  ^{N} s_i\mathbf{\psi}_i  + \frac{s_j}{a}(a \psi_j+b\psi_k) +  (s_k - \frac{bs_j}{a})\psi_k \\
& = & \mathbf{\Psi}'\mathbf{s}'
\end{IEEEeqnarray*}
By Eq.~(\ref{eq:suppunchanged}), we conclude that $\| \mathbf{s}'\|_0 = \|\mathbf{s}\|_0$, hence completes the proof.
\end{proof}
Obviously, the operator $\mathbb{F}_3(\cdot)$ can be applied to three or more columns as long as all of the chosen columns are either in $\supp(\mathbf{s})$
or not. Finally, we provide an example to further illustrate Property~\ref{pro:basischange3}.
The grayscale image ``Lena" with size $512 \times 512$, as shown in Fig~\ref{fig3:DCT2Energy}a), is transformed using
RPFrCT with orders $\alpha = 0.99$ and $\beta= 0.95$.
Figure~\ref{fig3:DCT2Energy}b) shows the absolute value of the RPFrCT coefficients under
the logarithm base. It is clear that the energy of the RPFrCT coefficients matrix is localized, specifically,
they are concentrated at the upper-left corner of the four sub-blocks. Thus, we
can apply Property~\ref{pro:basischange3} to the RPFrCT basis $\mathbf{\Psi} = (\mathbf{R}_{\beta} \otimes \mathbf{R}_{\alpha})^T$ accordingly. {A similar effect can be observed in the parameterized DWT and DCT settings.}

\section{{Discussions and Security Analysis}}
\label{sec:discussion}
{
We have demonstrated the possibility of using BLP-CS as a joint data acquisition and protection model for MTS purpose. This section aims to compare the basic OTS CS cipher and BLP-CS cipher from the viewpoints of complexity and security.}

{
\subsection{Complexity}
\label{subsec:complexity}
Suppose we have constructed a RPFrCT matrix $\mathbf{R}_{\alpha}$ with appropriate fractional order $\alpha$, a $M\times 1$ signal $\mathbf{x}$ can be sparsified by $\mathbf{R}_{\alpha}\mathbf{x} =  \mathbf{s}$. All the techniques on manipulating the sparsifying basis $\mathbf{R}_{\alpha}^T$ introduced in Sec.~\ref{sec:TypeII} can be unified to the following matrix notation\footnote{We are aware of the fact that any parameterized orthonormal transform with good sparsifying capability can play the role of $\mathbf{R}_{\alpha}^T$.}, i.e.,
\begin{IEEEeqnarray}{rCl}
\mathbf{\Psi}_{K}  & = & \mathbf{R}_{\alpha}^T  \mathbf{P} \mathbf{D} \mathbf{Q} \nonumber,
\end{IEEEeqnarray}
where $\mathbf{D}$, $\mathbf{P}$ and $\mathbf{Q}$ are matrices determined by operators $\mathbb{F}_1$, $\mathbb{F}_2$ and $\mathbb{F}_3$, respectively. It worth mentioning that $\mathbf{x}=\mathbf{\Psi}_{K} \mathbf{s'} = \mathbf{R}_{\alpha}^T \mathbf{s}$ with $\|\mathbf{s'}\|_0 = \|\mathbf{s}\|_0$. Recall from Sec.~\ref{subsec:newmodel}, the encoding of BLP-CS is governed by
\begin{IEEEeqnarray}{rCl}
\label{eq:blpsampling}
 \mathbf{y} = \mathbf{\Phi}\mathbf{x} &=& \mathbf{A}_K \mathbf{\Psi}_{K}^{-1} \mathbf{x},
\end{IEEEeqnarray}
and the decoding should follow a two-step reconstruction, i.e.,
\begin{IEEEeqnarray}{rCl}
  \min \|\mathbf{{s'}}\|_1~~\text{subject to }\mathbf{y} &=&  \mathbf{\Phi}  \mathbf{x}=\mathbf{A}_K  \mathbf{{s'}}, \nonumber \\
  \mathbf{x} &=& \mathbf{\Psi}_K \mathbf{s'}.
\label{eq:blpdecoding}
\end{IEEEeqnarray}
Once a well-designed key schedule is given\footnote{The design of an effective key scheduling process is not considered in this paper since our concern is only the secrecy of CS paradigm. We also note that this is a common treatment for all the state-of-the-art works on this topic.}, a trusted third party 
can produce $\mathbf{\Phi}$, $\mathbf{A}_K$ and $\mathbf{\Psi}_K$ faithfully and transmit them to the encoder and decoder. An alternative option is that the encoder and decoder produce their own matrix key on the air using the agreed key schedule from the same root key. We assume the OTS CS model also adopts the same matrix key generation process for a fair comparison.
}

{
We first take a look at the encoder side. For the former situation, where the matrix key is produced by the trusted party and then delivered to both the CS encoder and decoder, the encoding complexity of the BLP-CS model outperforms that of the OTS CS model since it does not bring extra communication cost once the key is set.
For the later situation, the
encoding complexity of the OTS CS model is lower than that of the BLP-CS model at the first glimpse due to the reason that the encoding process of the second model involves a matrix multiplication, i.e., $\mathbf{A}_K \mathbf{\Psi}_{K}^{-1}$, in the key generation process. Nevertheless, since the OTS CS system requires updating the measurement matrix in every sampling, the BLP-CS model outperforms OTS CS after sampling $(2f'+f)/f'$ times. Here, $f$ and $f'$ refer to the complexity of the matrix multiplication and the matrix key generation, respectively.
}

{
At the decoder side, the Moore-Penrose pseudoinverse of the sensing matrix $\mathbf{A}_K$ need to be calculated in every iteration of some $l_1$ optimization algorithms \cite{Boyd:Convex:08}, for example, orthogonal matching pursuit \cite{tropp2007signal}. The complexity of this operation dominates the overall complexity in CS reconstruction. As such, if some off-line techniques can be employed to calculate the pseudoinverse of $\mathbf{A}_K$, the complexity of the reconstruction can be largely reduced. For the OTS CS system, this is impossible since the measurement matrix is never re-used.
}
{
\subsection{Security}
\begin{enumerate}[I.]
\item{\textit{Brute-force and Ciphertext-only Attacks}} \\
We employ the existing results presented in \cite{Rachlin:secrecy:08,cambareri2015low} to show that the BLP-CS preserves most secrecy features of the OTS CS-based cipher under these two attacks.
\begin{Theorem} \cite[Theorem~1 and Corollary~1]{Rachlin:secrecy:08}
\label{theorem:wrongrec}
Let $\mathbf{A}$ and $\mathbf{A}'$ be $K\times M$ Gaussian matrices. Let $\mathbf{x}$ be $k$-sparse with respect to the canonic basis and $\mathbf{y}=\mathbf{A}\mathbf{x}$. If $K>k$, then $l_0$ problem (\ref{eq:l0hard}) and $l_1$ problem (\ref{eq:l1opt}) will yield an $K$-sparse solution $\mathbf{x}'$ with probability one such that $\mathbf{y}=\mathbf{A}'\mathbf{x}'$.
\end{Theorem}
We first examine the case of brute-force attack, i.e., the attacker try to guess possible measurement matrices and use them for decoding.
 Referring to Theorem~\ref{theorem:wrongrec}, the $l_0$ or $l_1$ recovery governed by a wrong sensing matrix $\mathbf{A}_K$ will lead to an incorrect reconstruction with probability one. Thus the OTS CS-based cipher can guarantee computational secrecy if the key space is large enough to make systematic search of all the keys (sensing matrices) impossible. This result can be directly applied to our BLP-CS model. According Eqs.~(\ref{eq:blpsampling}) and (\ref{eq:blpdecoding}), we can conclude that BLP-CS is computationally strong even if the attacker can successfully retrieved the secret sparsifying basis $\mathbf{\Psi}_K$. In this concern, the transform encryption approach enhances the security level of the basic CS paradigm. \\
An interesting security feature of the OTS CS cryptosystem  under ciphertext-only attack is the asymptotic spherical secrecy \cite{cambareri2015low}. This type of secrecy states that any two different plaintexts (sparse signals to be sampled in this context) with equal power 
remain approximately indistinguishable from their measurement vectors when CS operates under the RIP framework. Alternatively, we can intercept this property as only the energy of the measurements carries information about the signal. A bird's-eye view of why this asymptotic spherical secrecy holds for the OTS CS cipher may refer to the definition of RIP, which states that the CS encoding should obey an energy-preserving guarantee. A theoretical proof about this property can be found in \cite{cambareri2015low}.\\
As we demonstrated in Eqs.~(\ref{eq:blpsampling}) and (\ref{eq:blpdecoding}), the proposed BLP-CS model works under the seemingly RIPless theory if one cannot determine $\mathbf{A}_K$ and $\mathbf{\Psi}_K$. Therefore, the energy-preserving constraint introduced by RIP is unapplicable to this setting. As such, we can conclude that the measurements (ciphertext) carries no information about the signal (plaintext) when a single ciphertext is observed. The BLP-CS and the OTS CS ciphers have the following major difference: when multiple ciphertexts are observed by the attacker, he is aware of the fact that two plaintexts must be similar if their corresponding ciphertexts are close to each other in the Euclidean space. This is caused by the multi-time usage of the same measurement matrix and the linear encoder. Surely the OTS CS cipher is more secure then the BLP-CS cipher from this point of view.
Nevertheless, as mentioned in Sec.~\ref{sec:intro}, this is a favorable property that promotes the source coding gain from a system point-of-view \cite{mun2012dpcm}. This property also finds its way in privacy-preserving video surveillance systems \cite{tong2011compressive}:
assume the attacker happens to know some pairs of plaintext and ciphertext, such as static video scenes and their corresponding measurement vectors, and he want to retrieve privacy-sensitive data from a new intercepted ciphertext. After studying the Euclidean distance of the new ciphertext, he comes to realize that plaintext corresponding to the new ciphertext contains privacy-sensitive data. However, the decryption of this ciphertext requires full knowledge of the matrix key $\mathbf{A}_K$ and $\mathbf{\Phi}_K$. This leads to our discussion of resistance of the BLP-CS cipher with respect to plaintext attacks.
\item{\textit{Plaintext Attacks}} \\
As discussed in Sec.~\ref{sec:II}, the data complexity of retrieving a general measurement matrix (the secret key) is $M$ independent plaintexts and their corresponding ciphertexts in any basic CS-based cipher. If the used measurement matrix is Bernoulli, a single plaintext in the form
$\mathbf{x}=(2^0, 2^1, \cdots, 2^M)^T$ and the corresponding ciphertext can be utilized to recover the Bernoulli measurement matrix completely\footnote{One can imagine the role of a $\{+1, -1\}$ matrix as that of a $\{0, 1\}$ matrix, the proof can be found in \cite{cambareriknown}. A vector composed by $\{0, 1\}$ can be recovered from the inner product of this vector and $\mathbf{x}$. }.
Based on these knowledge, investigating the resistance of the OTS CS cryptosystem is a trivial work. We hereby focus on the BLP-CS cipher. Referring to Eq.~(\ref{eq:blpsampling}), the attacker can retrieve $\mathbf{\Phi}$ from $M$ independent plaintext-ciphertext pairs. By construction, $\mathbf{\Phi}$ is a non-RIP matrix. Thus the conclusion drawn from Theorem~\ref{theorem:RIPless} assures that a straightforward use $\mathbf{\Phi}$ in the $l_1$ optimization problem (\ref{eq:l1opt}) is not applicable. Considering that the $l_0$ optimization problem (\ref{eq:l0hard}) is NP-hard \cite{Tao:BPunderRIP:TIT05}, the attacker tries to decompose $\mathbf{\Phi}$ with the form $\mathbf{\Phi} = \mathbf{E}\mathbf{F}$, with the constraint that entries of $\mathbf{E}$ should observe certain kind of distribution (Gaussian or Bernoulli). In particular, $\mathbf{F}$ is the product of an elementary matrix and an orthonormal matrix. \\
If the decomposition is unique or the possible number of decompositions is very limited, i.e.,   polynomial function of $M$, the attacker can determine the matrix key $\mathbf{A}_K$ and $\mathbf{\Psi}_{K}^{-1}$ and the BLP-CS cryptosystem is regarded as fail to resist plaintext attacks. To summarize, we conclude that the number of decompositions should be at least $O(M!)$, thus making the search for the true one inconclusive\footnote{This is even worse than directly solving the NP-hard $l_0$ problem (\ref{eq:l0hard}), who has a complexity $M \choose k$.}. The conclusion is based on the simple fact $\mathbf{E}\mathbf{F} = (\mathbf{E}\mathbf{P})(\mathbf{P}^T\mathbf{F})$, where $\mathbf{P}$ is a $M \times M$ random permutation matrix. As we can see, distribution of all the entries of $(\mathbf{E}\mathbf{P})$ is exactly the same as that of $\mathbf{E}$ and $\mathbf{P}^T$ represents elementary row operation on $\mathbf{F}$. As such, the attacker cannot distinguish the decomposition result $\mathbf{E}$ and $\mathbf{F}$ from $(\mathbf{E}\mathbf{P})$ and $(\mathbf{P}^T\mathbf{F})$.
\end{enumerate}
}

\section{BLP-CS for Digital Images}
\label{sec:Sec4}
In this section, the proposed {BLP-CS model is applied as a joint data acquisition and protection layer for digital images. The aim is to provide an intuitive interpretation of how a cryptographic random scrambling can relax RIP of the measurement matrix and substantially reduce the decoding complexity, i.e., parallel reconstruction. Moreover, some other features owned by a basic CS paradigm, such as robust to packet loss and noise, are also observed.}

{
We now consider a $2$D image $\mathbf{X}$ with $M= n\times n$ pixels. If the chosen parameterized transform is RPFrCT, the basis for $\mathbf{X}$ is $(\mathbf{R}_{\beta}^T\otimes \mathbf{R}_{\alpha}^T)$ according to Eq.~(\ref{eq:2dbasis}).
Following the same approach adopted in \cite{duarte2008single}, the encoding stage can be written as
\begin{IEEEeqnarray}{rCl}
\vec(\mathbf{Y}) = [\mathbf{y}_1, \mathbf{y}_2, \cdots, \mathbf{y}_n]^T= \mathbf{\Phi}\vec(\mathbf{X}), \nonumber
\end{IEEEeqnarray}
where $\mathbf{\Phi}$ is the product of the $K\times M$ key-dependent sensing matrix $\mathbf{A}_K$ and the $M\times M$ key-dependent basis $\mathbf{\Psi}^{-1}_{K}$ having the form
\begin{IEEEeqnarray}{rCl}
\mathbf{\Psi}^{-1}_{K} = \mathbf{D}^{-1}\mathbf{P}^{T}(\mathbf{R}_{\beta}^T\otimes \mathbf{R}_{\alpha}^T), \nonumber
\end{IEEEeqnarray}
and
\begin{IEEEeqnarray}{rCl}
\mathbf{A}_K = \left[
\begin{array}{cccc}
\mathbf{A}_1   &                &               &\\
               & \mathbf{A}_2   &               &\\
               &                &\ddots         &\\
               &                &               &\mathbf{A}_n
\end{array} \nonumber \right]
\end{IEEEeqnarray}
with $\mathbf{A}_j =  \mathbf{A}$ for $j \in \{1, \cdots n\}$ being Gaussian matrices. As we discussed in Sec.~\ref{subsec:complexity}, repeatedly using the same sensing matrix for different signal segments can speed up the reconstruction if some off-line mechanism is allowed to calculate the pseudoinverse of $\mathbf{A}$ in advance.
}

\begin{figure}[!htb]
\centering
\begin{minipage}[t]{\imagewidthb}
\centering
\includegraphics[width=\imagewidthb]{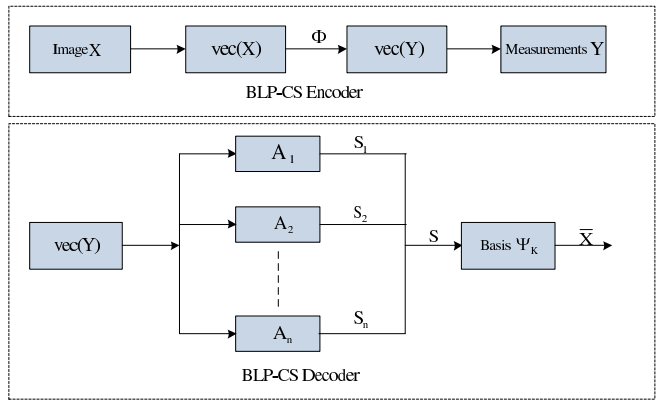}
\end{minipage}
\caption{Block diagram of BLP-CS for digital images.}
\label{fig:PCSRecons}
\end{figure}

{
According to Secs.~\ref{sec:TypeI} and \ref{sec:TypeII}, $\vec(\mathbf{S}) = [\mathbf{s}_1, \mathbf{s}_2, \cdots, \mathbf{s}_n]^T = \mathbf{\Psi}^{-1}_{K}\vec(\mathbf{X})$ is sparse in the canonical basis. Referring to property~\ref{pro:permutation} and Eq~(\ref{eq:PCSreconstruction}), a parallel construction is applied as
\begin{IEEEeqnarray}{rCl}
\label{eq:pcsrec}
 \min \|\mathbf{s}_j\|_1~~\text{subject to }\mathbf{y}_j &=&\mathbf{A} \mathbf{s}_j.
\end{IEEEeqnarray}
for all $j\in \{1, 2,\cdots, n\}$. Finally, the recovered image is given by $\vec(\mathbf{\bar{X}})=\mathbf{\Psi}_{K} \vec(\mathbf{S})$.
} A block diagram of the whole system is depicted in Fig.~\ref{fig:PCSRecons}.
{In summary, this system is a instance of the simplified BLP-CS model.}

{
To further illustrate how the random scrambling $\mathbf{P}$ relaxes the RIP requirement of the sensing matrix $\mathbf{A}$,
we consider another sampling configuration
\begin{IEEEeqnarray}{rCl}
\vec(\mathbf{Y}) = \mathbf{\Phi}\vec(\mathbf{X}), \nonumber
\end{IEEEeqnarray}
where $\mathbf{\Phi} =\mathbf{A}_K  \mathbf{\hat{\Psi}}^{-1}_{K}$ with $\mathbf{A}_K$ is the same as defined above and $\mathbf{\hat{\Psi}}^{-1}_{K}= \mathbf{D}^{-1}(\mathbf{R}_{\beta}^T\otimes \mathbf{R}_{\alpha}^T)$. Here, we note that the only difference of $\mathbf{{\Psi}}^{-1}_{K}$ and $\mathbf{\hat{\Psi}}^{-1}_{K}$ is the permutation matrix $\mathbf{P}$. The reconstruction is exactly the same as that of Eq.~(\ref{eq:pcsrec}). By construction, this is a special form of block-based compressive sampling (BCS) \cite{Gan:BCS:07}, where each block is a column of the frequency coefficients, together with block independent recovery. We call this model BCS-In. We also note that using the smoothed projected Landweber operator can largely improve the BCS reconstrution quality at relatively low extra computation overhead \cite{fowler2011multiscale}. However, the study of embedding the smoothed projected Landweber operator in the BLP-CS reconstruction is out of the scope of this paper.
}

{
Four representative images, ``Lena", ``Peppers", ``Cameraman" and ``Baboon" of size $512\times 512$ are used as our test images. The tests are carried out under different sampling rate SR $= \frac{K}{M}\times 100\%$. The reconstruction quality is evaluated in terms of
average\footnote{$\mathbb{E}$ denotes calculate average over $100$ tests.} peak signal-to-noise ratio, APSNR (dB) $=10 \cdot \log_{10}\mathbb{E} \left( \frac{M255^2}{\|\vec(\mathbf{X})- \vec(\mathbf{\bar{X}})\|_2^2}\right)$. The results are listed in Table~\ref{table:PSNRcomparision}
and they support the conclusion of property~\ref{pro:permutation}, i.e., a cryptographic random scrambling helps make the column sparsity level of $\mathbf{S}$ uniform.
The last point worth mentioning is that random scrambling is suitable for all kind of $2$D sparse data (all kind of sparsifying coefficients under parameterized orthonormal transform), which extends the result that zig-zag scrambling works for DCT$2$ coefficients \cite{Fang:permutation:TSP13}.
}
\begin{table*}[!htbp]
\centering
\caption{Comparison between {BLP-CS and BCS-In} in terms of {A}PSNR at different SRs.}
\begin{tabular}{c|c|c|c|c|c|c|c|c}
\hline
{SR}           & \multicolumn{2}{c|}{10\%}  & \multicolumn{2}{c|}{30\%}  & \multicolumn{2}{c|}{50\%}  &\multicolumn{2}{c}{70\%}  \\ \cline{1-9}
{Model}        & {BLP-CS} &{BCS-In}              & {BLP-CS} &{BCS-In}              & {BLP-CS}          & {BCS-In}    &{BLP-CS}     &{BCS-In}         \\ 
\hline \hline
{``Lena"}      &$21.6$      &$15.5$        &$27.5$        &$23.3$            &$31.4$       &$27.3$   &$35.7$  &$32.1$\\
\hline
{``Peppers"}   &$20.9$      &$14.4$        &$27.2$        &$22.6$            &$30.9$       &$27.9$   &$34.7$  &$32.5$\\
\hline
{``Cameraman"} &$19.2$      &$13.0$        &$24.8$        &$21.5$            &$28.6$       &$27.4$   &$32.9$  &$32.8$\\
\hline
{``Baboon"}    &$17.8$      &$9.7$         &$20.2$        &$17.6$            &$22.6$       &$21.3$   &$25.8$  &$25.2$\\
\hline
\end{tabular}
\label{table:PSNRcomparision}
\end{table*}

{The basic CS paradigm that works under RIP theory is known to be robust}
with respect to transmission imperfections such as noise or packet loss \cite{dimakis2009lp,laska2011democracy}. {
Since the new proposal works under the RIPless theory at only the encoder but RIP theory at the decoder, we expect the same property in our approach.} To  quantitatively study this, we evaluate the robustness of the proposed framework with respect to additive white Gaussian noise (AWGN) and various packet loss rates (PLRs). In the former case, we artificially add a zero-mean normal distribution random sequence with variance $1$  to the measurements while in the latter we randomly discard certain number of measurements governed by PLR. Then we {perform reconstruction on} the corrupted measurements. In real applications, PLR can be up to $30\%$ \cite{Zhao:PacketLoss:03ACM} and we measure the quality of the reconstruction in terms of APSNR at $10\%$, $20\%$ and $30\%$ PLR, respectively. These tests were carried out using the ``Lena" image, but similar results were obtained using other images. As observed from Table~\ref{table:NoisePsnr}, our scheme is almost immune to AWGN when we compare the APSNR of the ideal {case} and the one with AWGN. In addition, comparing the APSNRs at different levels of PLR, we found that the reduction rate of APSNR is linear to the increasing rate of PLR, which implies that all measurements are of the same importance \cite{laska2011democracy}.
\begin{table}[!htbp]
\centering
\caption{APSNR of the reconstructions under AWGN and various PLRs.}
\begin{tabular}{c|c|c|c|c}
\hline
{SR}           & {0.1}  & {0.3}  & {0.5}  &{0.7}  \\ \cline{1-2}
\hline \hline
{Ideal BLP-CS}          &21.6     &27.5       &31.4       &35.7    \\
\hline
{BLP-CS AWGN}          &21.8     &27.4       &31.3       &34.9          \\
\hline
{BLP-CS~$10\%$ PLR }   &21.7     &26.8       &30.5       &34.1          \\
\hline
{BLP-CS~$20\%$ PLR }   &20.9     &26.2       &29.5       &32.7          \\
\hline
{BLP-CS~$30\%$ PLR }   &19.9     &25.5       &28.5       &31.3          \\
\hline
\end{tabular}
\label{table:NoisePsnr}
\end{table}

\section{Conclusion}
\label{sec:conclusion}
{
To realize the MTS usage of CS cryptosystem,
some approaches have already been proposed. Typical examples include scrambling in different domains \cite{zeng2012scrambling,huang2015compression,yushu:SCS:14arxiv,wu2014low} and cascading the DRPE technique \cite{deepan2014multiple,rawat2015compressive,li2015compressive}.
However, we have shown that they fail to satisfy the security requirement. In this concern, we suggest a BLP-CS model by making use of the non-RIP measurement matrix construction.
Our approach differs from existing ones in two aspects: 1)~the RIPless CS theory is firstly applied for providing the security features of a CS-based cipher; 2)~the role of the sparsifying basis for the secrecy of CS is revealed.
}

{The security of the BLP-CS model is discussed from various aspects, such as brute-force attack, ciphertext-only attack and plaintext attacks. Special attention has been paid to the plaintext attacks since it is widely accepted that basic CS model is immune to brute-force attack and ciphertext-only attack \cite{Rachlin:secrecy:08,cambareri2015low}.
Under plaintext attacks, we have demonstrated that the number of candidate sensing matrices and sparsifying basis matrices that match the information inferred by the attacker is huge. Therefore, the searching of the true sensing matrix and sparsifying basis matrix is impossible.
}

{
Finally, we apply the proposed model for the purpose of secure compressive image sampling. Both theoretical analyses and experimental results support our expectation, i.e., random scrambling plays a critical role in relaxing the RIP requirement of the measurement matrix and flavoring a PCS reconstruction for $2$D sparse signals. Other features of a basic CS system, such as robust to packet loss and noise, are also observed.
}


\small{
\bibliographystyle{IEEEtran}
\bibliography{SecurePCS_8_3}}

\end{document}